\documentclass[twocolumn,floatfix]{revtex4}
\usepackage{graphicx}
\usepackage{amsmath}
\usepackage{amssymb}
\usepackage{bm}

\begin{document}

\title{Annihilation of colliding Bogoliubov quasiparticles reveals their Majorana nature}
\author{C. W. J. Beenakker}
\affiliation{Instituut-Lorentz, Universiteit Leiden, P.O. Box 9506, 2300 RA Leiden, The Netherlands}
\date{December 2013}
\begin{abstract}
The single-particle excitations of a superconductor are coherent superpositions of electrons and holes near the Fermi level, called Bogoliubov quasiparticles. They are Majorana fermions, meaning that pairs of quasiparticles can annihilate. We calculate the annihilation probability at a beam splitter for chiral quantum Hall edge states, obtaining a $1\pm\cos\phi$ dependence on the phase difference $\phi$ of the superconductors from which the excitations originated (with the $\pm$ sign distinguishing singlet and triplet pairing). This provides for a nonlocal measurement of the superconducting phase in the absence of any supercurrent. 
\end{abstract}
\maketitle

Condensed matter analogies of concepts from particle physics are a source of much inspiration, and many of these involve superconductors or superfluids \cite{Vol03}. Majorana's old idea \cite{Maj37} that a spin-$1/2$ particle (such as a neutrino) might be its own antiparticle has returned \cite{Wil09} in the context of low-dimensional superconductors, inspiring an intense theoretical and experimental search for condensed matter realizations of Majorana fermions \cite{Fra13}. The search has concentrated on Majorana zero-modes \cite{Ali12,Sta13,Bee13} --- midgap states (at the Fermi level $E=0$) bound to a defect in a superconductor with broken spin-rotation and time-reversal symmetry (a so-called topological superconductor \cite{Has10,Qi11}). The name Majorana {\em zero-mode} (or {\em Majorino} \cite{Wil14}) is preferred over Majorana {\em fermion}, since they are not fermions at all but have a non-Abelian exchange statistics \cite{Rea00}.

Majorana {\em fermions}, in the original sense of the word, do exist in superconductors, in fact they are ubiquitous: The time-dependent four-component Bogoliubov-De Gennes wave equation for quasiparticle excitations (so-called Bogoliubov quasiparticles) can be brought to a real form by a $4\times 4$ unitary transformation ${\cal U}$ \cite{Sen00}, in direct analogy to the real Eddington-Majorana wave equation of particle physics \cite{Maj37,Edi28}. A real wave equation implies the linear relation $\Psi^{\dagger}(\bm{r},t)={\cal U}\Psi(\bm{r},t)$ between the particle and antiparticle field operators, which is the hallmark of a Majorana fermion. As argued forcefully by Chamon et al.\ \cite{Cha10}, fermionic statistics plus superconductivity by itself produces Majorana fermions, irrespective of considerations of dimensionality, topology, or broken symmetries.

Here we propose an experiment to probe the Majorana nature of Bogoliubov quasiparticles in conventional, nontopological, superconductors. Existing proposals apply to topological superconductors \cite{Fu09,Akh09,Law09,Chu11,Str11,Li12,Die13,Hou13,Wie13,Alo13,Par13}, where Majorana fermions appear as charge-neutral edge states with a distinct signature in {\sc dc} transport experiments. In contrast, the Bogoliubov quasiparticles of a nontopological superconductor have charge expectation value $\bar{q}\neq 0$, so their Majorana nature remains hidden in the energy domain probed by {\sc dc} transport. 

It is in the \textit{time domain} that the wave equation takes on a real form and that particle and antiparticle operators are linearly related. We will show that the Majorana relation manifests itself in high-frequency shot noise correlators, that can detect the annihilation of a pair of Bogoliubov quasiparticles originating from two identical superconductors (differing only in their phase). These quasiparticles can annihilate for nonzero $\bar{q}$ because of quantum fluctuations of the charge (with variance ${\rm var}\,q$). We calculate the annihilation probability ${\cal P}$ and find that it oscillates with the phase difference $\phi$,
\begin{equation}
{\cal P}=\tfrac{1}{2}(1+\cos\phi)\,{\rm var}\,(q/e).\label{calPdef}
\end{equation}
This could provide a way to detect the nonlocal Josephson effect \cite{Iaz10}, existing in the absence of any supercurrent flowing between the superconductors.

\begin{figure}[tb]
\centerline{\includegraphics[width=0.8\linewidth]{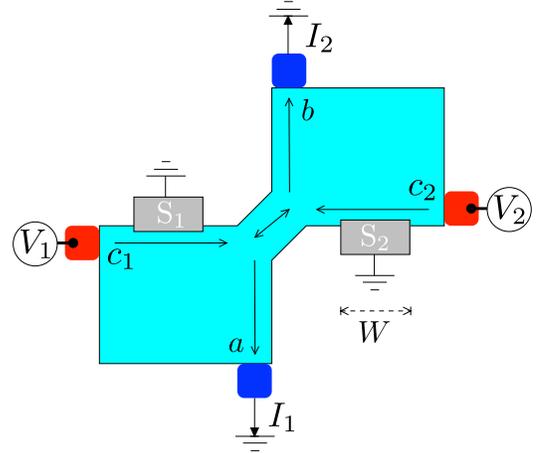}}
\caption{Two-particle interferometer for Bogoliubov quasiparticles. Shown is a 2D electron gas in a perpendicular magnetic field (light blue), with chiral edge channels at the edges (arrows indicate the direction of motion and $a,b,c_p$ denote the quasiparticle operators). A constriction at the center forms a beam splitter. Current is injected at the two ends (red), biased at voltages $V_1$ and $V_2$. Upon passing along a superconducting electrode (grey, labeled S), repeated Andreev reflection converts the electrons into a coherent superposition of electrons and holes. The collision and pairwise annihilation of these Bogoliubov quasiparticles is detected by correlating the {\sc ac} currents $I_{1}$ and $I_{2}$. 
}
\label{fig_beamsplitter}
\end{figure}

We consider the beam splitter geometry of Fig.\ \ref{fig_beamsplitter}, in which electrons are injected from two voltage sources at one side of the beam splitter and the fluctuating currents $I_1(t)$ and $I_2(t)$ are correlated at the other side at microwave frequencies $\omega>0$,
\begin{equation}
P(\omega)=\int_{-\infty}^{\infty}dt\,e^{i\omega t}\langle I_{1}(0)I_{2}(t)\rangle.\label{Pomegadef}
\end{equation}
Such two-particle interferometers have been implemented using the quantum Hall edge channels of a two-dimensional (2D) electron gas as chiral (uni-directional) wave guides, to realize the electronic analogues of the Hanbury-Brown-Twiss (HBT) experiment \cite{Hen99,Oli99,Ned07} and the Hong-Ou-Mandel (HOM) experiment \cite{Boc13,Dub13}. 

The setup we propose here differs in one essential aspect: Before reaching the beam splitter, the electrons are partially Andreev reflected at a superconducting electrode. Andreev reflection in the quantum Hall effect regime has been reported in InAs quantum wells \cite{Ero05,Bat07} and in graphene monolayers \cite{Pop12,Kom12,Ric12}. In graphene, which has small spin-orbit coupling, the Andreev reflected hole is in the opposite spin band as the electron (spin-singlet pairing). The strong spin-orbit coupling in InAs permits spin-triplet pairing (electron and hole in the same spin band).

We contrast these two cases by taking a twofold spin-degenerate edge channel for spin-singlet pairing and one single spin-polarized edge channel for spin-triplet pairing. For spin-singlet pairing we therefore need a vector of four annihilation operators $a=(a_{e\uparrow},a_{e\downarrow},a_{h\uparrow},a_{h\downarrow})=\{a_{\tau\sigma}\}$, to accomodate electrons and holes ($\tau=e,h$) in both spin bands ($\sigma=\uparrow,\downarrow$), while for spin-triplet pairing the two operators $a=(a_{e\uparrow},a_{h\uparrow})$ suffice. The creation and annihilation operators of these Bogoliubov quasiparticles are related by particle-hole symmetry \cite{note1}, 
\begin{equation}
a(E)=\tau_{x}a^{\dagger}(-E).\label{adaggerarelation}
\end{equation}
The Pauli matrices $\tau_{i}$ and $\sigma_{i}$ act, respectively on the electron-hole and spin degree of freedom. The anticommutation relations thus have an unusual form,
\begin{subequations}
\label{anticommutation}
\begin{align}
&\{a_{\tau\sigma}^{\vphantom{\dagger}}(E),a_{\tau'\sigma'}^{\dagger}(E')\}=\delta(E-E')\delta_{\tau\tau'}\delta_{\sigma\sigma'},\label{anticommutationa}\\
&\{a_{\tau\sigma}(E),a_{\tau'\sigma'}(E')\}=\nonumber\\
&\quad
\begin{cases}
\delta(E+E')\delta_{\sigma\sigma'}&\text{if $\tau,\tau'$ is $e,h$ or $h,e$}\\
0&{\rm otherwise}.
\end{cases}
\label{anticommutationb}
\end{align}
\end{subequations}
The nonzero anticommutator of two annihilation operators is the hallmark of a Majorana fermion \cite{Cha10}. 

The electrical current operator is represented by
\begin{equation}
I(t)=e a^{\dagger}(t)\tau_{z}a(t),\;\;a(t)=\frac{1}{\sqrt{4\pi}}\int_{-\infty}^{\infty}dE\,e^{-iEt}a(E).\label{Itdef}
\end{equation}
The Pauli matrix $\tau_z$ accounts for the opposite charge of electron and hole. (For notational convenience we set $\hbar=1$ and take the electron charge $e>0$.) To distinguish the currents $I_1$ and $I_2$, we will denote the quasiparticle operators at contact $2$ by $a$ and those at contact $1$ by $b$. The current correlator \eqref{Pomegadef} then takes the form
\begin{align}
P(\omega)={}&\tfrac{1}{4}G_{0}\int_{-\infty}^{\infty}dE\int_{-\infty}^{\infty}dE'\int_{-\infty}^{\infty}dE''\nonumber\\
&\times\bigl\langle b^{\dagger}(E')\tau_{z}b(E'')a^{\dagger}(E-\omega)\tau_{z}a(E)\bigr\rangle,\label{P12ab}
\end{align}
with $G_{0}=e^2/h$ the conductance quantum.

Using the Majorana relation \eqref{adaggerarelation} we can rewrite Eq.\ \eqref{P12ab} so that only positive energies appear \cite{note2}. Only products of an equal number of creation and annihilation operators contribute, resulting in
\begin{align}
&P(\omega)=G_{0}\int_{0}^{\infty}dE\int_{0}^{\infty}dE'\int_{0}^{\infty}dE''\nonumber\\
&\quad\mbox{}\times\bigl\langle b^{\dagger}(E')\tau_{z}b(E'')a^{\dagger}(E)\tau_{z}a(E+\omega)\nonumber\\
&+\tfrac{1}{4}\theta(\omega-E)b^{\dagger}(E')\tau_{y}b^{\dagger}(E'')a(\omega-E)\tau_{y}a(E)\bigr\rangle,\label{noisepower2}
\end{align}
with $\theta(x)$ the unit step function. Both terms describe an inelastic process accompanied by the emission of a photon at frequency $\omega$. The difference is that the term with $\tau_z$ is a single-particle process (relaxation of a quasiparticle from energy $E+\omega\mapsto E$), while the term with $\tau_y$ is a two-particle process (pairwise annihilation of quasiparticles at energy $E$ and $\omega-E$). The appearance of this last term is a direct consequence of the Majorana relation \eqref{adaggerarelation}, which transforms $a^{\dagger}(E-\omega)\tau_{z}a(E)\mapsto a(\omega-E)\tau_{x}\tau_{z}a(E)$.

The quasiparticle operators $c_p$ injected towards the beam splitter by voltage contact $p=1,2$ are related to their counterparts $a,b$ behind the beam splitter by a scattering matrix. Since the voltage contacts are in local equilibrium, the expectation value of the $c_p$ operators is known, and in this way one obtains an expression for the noise correlator in terms of scattering matrix elements --- an approach pioneered by B\"{u}ttiker \cite{But91} and used recently to describe the electronic HBT and HOM experiments \cite{Sam04,Par12,Jon12}. 

Our new ingredient is the effect of the superconductor on the injected electrons. Propagation of the edge channel along the superconductor transforms the quasiparticle operators $c_{p}(E)\mapsto M_p(E)c_{p}(E)$ through a unitary transfer matrix constrained by particle-hole symmetry,
\begin{equation}
M_{p}(E)=\tau_{x}M_{p}^{\ast}(-E)\tau_{x}.\label{Mehsym}
\end{equation}
The effect of the beam splitter is described by the unitary transformation
\begin{equation}
\begin{split}
a=\sqrt{R}\,M_{1}c_{1}+\sqrt{1-R}\,M_{2}c_{2},\\
b=\sqrt{1-R}\,M_{1}c_{1}-\sqrt{R}\,M_{2}c_{2}.
\end{split}
\label{abc1c2relation}
\end{equation}
(For simplicity, we take an energy independent reflection probability $R$.) The voltage contacts inject quasiparticles in local equilibrium at temperature $T$ and chemical potential $eV>0$ (the same at both contacts), corresponding to expectation values
\begin{equation}
\langle c_{p,\tau\sigma}^{\dagger}(E)c_{q,\tau'\sigma'}(E')\rangle=\delta_{\tau\tau'}\delta_{\sigma\sigma'}\delta_{pq}\delta(E-E')f_{\tau}(E),\label{cpaaverage}
\end{equation}
with electron and hole Fermi functions,
\begin{equation}
f_{e}(E)=\frac{1}{1+e^{(E-eV)/k_{\rm B}T}},\;\;f_{h}(E)=1-f_{e}(-E).\label{fehdef}
\end{equation}
In what follows we focus on the low-temperature regime $k_{\rm B}T\ll eV$, when only electrons are injected by the voltage contacts: $f_{e}(E)=\theta(eV-E)\equiv f(E)$, while $f_{h}(E)=0$ for $E,V>0$.

The full correlator \eqref{noisepower2} decomposes into four terms,
\begin{align}
&P(\omega)=P_{11}(\omega)+P_{22}(\omega)+P_{12}(\omega)+P_{21}(\omega),\label{Pfourterms}\\
&P_{pq}(\omega)=-G_{0} R(1-R)(-1)^{p+q}\int_{0}^{\infty}dE\,f(E)\nonumber\\
&\mbox{}\times{\rm Tr}\,\bigl[ f(E+\omega)Z_{pq}(E+\omega,E)Z_{qp}(E,E+\omega)\nonumber\\
&\;\mbox{}+\tfrac{1}{2}\theta(\omega-E)f(\omega-E)Y_{pq}^{\ast}(E,\omega-E)Y_{qp}^{\vphantom{\ast}}(\omega-E,E)\bigr],\label{noisepower3}\\
&Z_{pq}(E,E')=\tfrac{1}{4}(1+\tau_{z})M_{p}^{\dagger}(E)\tau_{z}M_{q}^{\vphantom{\dagger}}(E')(1+\tau_{z}),\label{Zdef}\\
&Y_{pq}(E,E')=\tfrac{1}{4}(1+\tau_{z})M_{p}^{\rm T}(E)\tau_{y}M_{q}^{\vphantom{\rm T}}(E')(1+\tau_{z}).\label{Ydef}
\end{align}
The partial correlators $P_{11}$ and $P_{22}$ can be measured separately by biasing only voltage contact $1$ or $2$, respectively. The terms $P_{12}$ and $P_{21}$ describe the collision at the beam splitter of particles injected from contacts $1$ and $2$.

Transfer matrices of quantum Hall edge channels propagating along a superconducting contact (so-called Andreev edge channels) have been calculated in Ref.\ \cite{Ost11}. Their general form is constrained by unitarity and by the electron-hole symmetry relation \eqref{Mehsym}. A single spin-degenerate Andreev edge channel has transfer matrix
\begin{align}
&M_{p}=e^{iEt_{p}}e^{i\gamma_{p}\tau_{z}}U(\alpha_{p},\phi_{p},\beta_{p})e^{i\gamma'_{p}\tau_{z}},\label{Mpsinglet}\\
&U(\alpha,\phi,\beta)=\exp\bigl[i\alpha\sigma_{y}\otimes(\tau_{x}\cos\phi+\tau_{y}\sin\phi)+i\beta\tau_{z}\bigr].\label{Udef}
\end{align}
The $\tau_{z}$ terms account for relative phase shifts of electrons and holes in the magnetic field, while the terms $\sigma_{y}\otimes\tau_{x}\cos\phi_p$ and $\sigma_{y}\otimes\tau_{y}\sin\phi_p$ describe the electron-hole mixing by a spin-singlet pair potential with phase $\phi_{p}$. For a superconducting interface of width $W$ one has $\alpha\simeq W/l_{\rm S}$ and $\beta\simeq W/l_{\rm m}$, with $l_{\rm m}=(\hbar/eB)^{1/2}$ the magnetic length and $l_{\rm S}=\hbar v_{\rm edge}/\Delta$ the superconducting coherence length (for induced gap $\Delta$ and edge velocity $v_{\rm edge}$).

The presence of a $\tau_z$ term in the electron-hole rotation matrix \eqref{Udef} is inconvenient. With some algebra, it can be eliminated, resulting in
\begin{align}
&M_{p}=e^{iEt_{p}}e^{i(\gamma_{p}+\delta\gamma_{p})\tau_{z}}U(\bar{\alpha}_{p},\phi_{p},0)e^{i(\gamma'_{p}+\delta\gamma_{p})\tau_{z}},\label{Mpsinglet2}\\
&\sin\bar{\alpha}_{p}=(\alpha_{p}/\xi_{p})\sin\xi_{p},\;\;\tan 2\delta\gamma_{p}=(\beta_{p}/\xi_{p})\tan\xi_{p}.\label{baralphadef}
\end{align}
We have abbreviated $\xi_{p}=(\alpha_{p}^{2}+\beta_{p}^{2})^{1/2}$. Typically one has $l_{\rm m}\lesssim l_{\rm S}$, which implies $\bar{\alpha}\simeq(l_{\rm m}/l_{\rm S})\sin(W/l_{\rm m})$ and $\delta\gamma\simeq W/2l_{\rm m}$.

Substitution into Eq.\ \eqref{noisepower3} gives the partial correlators
\begin{align}
&P_{pq}(\omega)=-G_{0} R(1-R)(-1)^{p+q}\int_{0}^{\infty}dE\,f(E)\nonumber\\
&\mbox{}\times\bigl[ f(E+\omega)(g_{pq}+1)+\tfrac{1}{2}\theta(\omega-E)f(\omega-E)(g_{pq}-1)\bigr],\label{noisepower4}\\
&g_{pq}=\cos 2\bar{\alpha}_{p}\cos 2\bar{\alpha}_{q}-\cos\phi_{pq}\sin 2\bar{\alpha}_{p}\sin 2\bar{\alpha}_{q},\label{gpqdef}\\
&\phi_{pq}=\phi_{p}-\phi_{q}-2(\gamma_{p}+\delta\gamma_{p}-\gamma_{q}-\delta\gamma_{q}).\label{phipqdef}
\end{align}
The phase $\phi_{12}$ represents the gauge invariant phase difference between the two superconductors. Substituting $f(E)=\theta(eV-E)$, the integral over energy evaluates to
\begin{align}
&P_{pq}(\omega)=-G_{0} R(1-R)(-1)^{p+q}\nonumber\\
&\quad\mbox{}\times\bigl[2\Theta(eV-\omega)+\tfrac{1}{2}\Theta(2eV-\omega)(g_{pq}-1)\bigr],\label{P12result}
\end{align}
where we have defined the function $\Theta(x)=x\,\theta(x)$. Substitution into Eq.\ \eqref{Pfourterms} then gives the full correlator
\begin{equation}
P(\omega)=-\tfrac{1}{2}G_{0} R(1-R)\Theta(2eV-\omega)\bigl(g_{11}+g_{22}-2g_{12}\bigr).\label{Presult}
\end{equation}

\begin{figure}[tb]
\centerline{\includegraphics[width=0.8\linewidth]{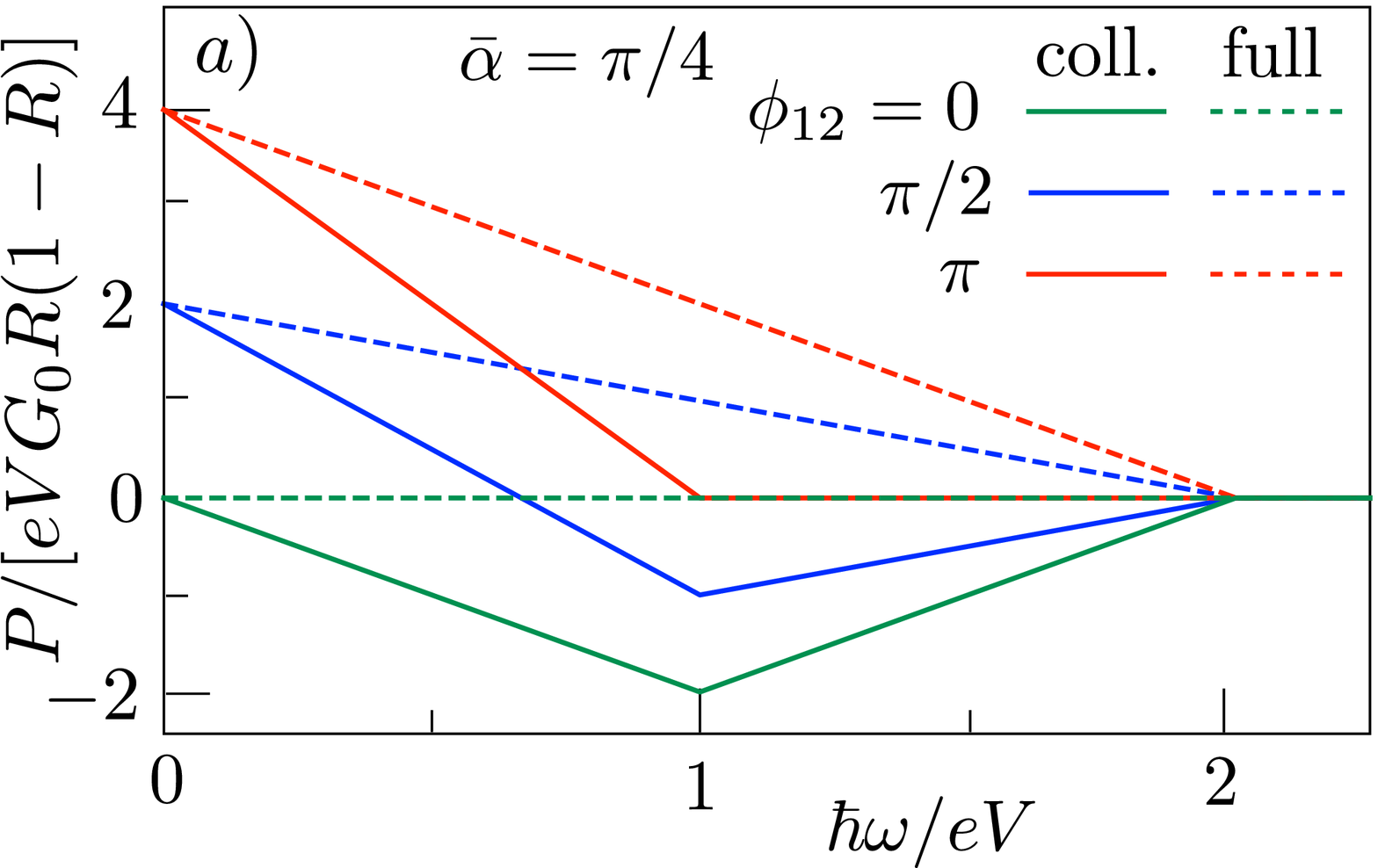}}
\centerline{\includegraphics[width=0.8\linewidth]{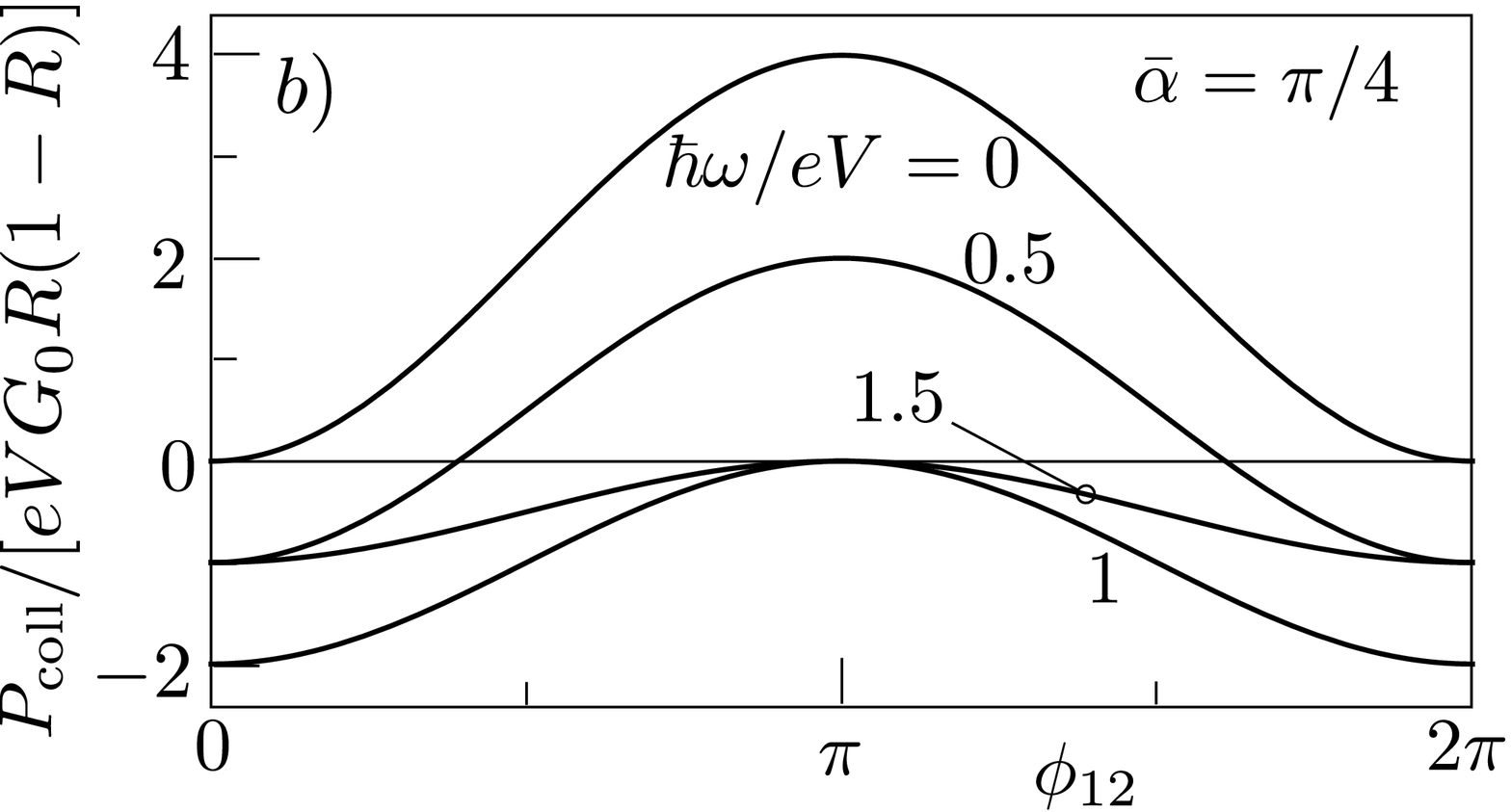}}
\caption{Noise correlator as a function of frequency (\textit{a}) and as a function of superconducting phase difference (\textit{b}). Panel \textit{a} shows both the collision term $P_{\rm coll}=2P_{12}$ and the full correlator $P_{\rm full}=P_{\rm coll}+P_{11}+P_{22}$, while panel \textit{b} shows only $P_{\rm coll}$ (the full correlator differs by a phase-independent offset, such that $P_{\rm full}=0$ at $\phi_{12}=0$). The curves are calculated from the general result \eqref{P12result} for spin-singlet pairing, with parameters $\bar{\alpha}_1=\bar{\alpha}_2=\pi/4$.
}
\label{fig_noise}
\end{figure}

In Fig.\ \ref{fig_noise} we compare the collision term $P_{\rm coll}=2P_{12}$ and the full correlator $P_{\rm full}=2P_{12}+P_{11}+P_{22}$.  The linear voltage dependence of $P_{\rm full}$ shown in Fig.\ \ref{fig_noise}\textit{a}, with a singularity (discontinuous derivative) at $\omega=2eV$, is known from two-terminal normal-superconducting junctions \cite{Tor01}. The collision term has an additional singularity at $\omega=eV$, signaling the frequency beyond which only pairwise annihilation of Bogoliubov quasiparticles contributes to the noise.

Fig.\ \ref{fig_noise}\textit{b} shows the dependence on the superconducting phase difference of the collision term, for the case of two identical superconductors, $\bar{\alpha}_1=\bar{\alpha}_2\equiv\bar{\alpha}$. In the annihilation regime $eV<\omega<2eV$ the general formula \eqref{P12result} then simplifies to
\begin{equation}
P_{12}(\omega)=-\tfrac{1}{2}G_{0} R(1-R)(2eV-\omega)(1+\cos\phi_{12})\sin^{2}2\bar{\alpha}.\label{P12simple}
\end{equation}
The factor $\sin^{2}2\bar{\alpha}={\rm var}\,(q/e)$ is the variance of the quasiparticle charge, cf.\ Eq.\ \eqref{calPdef}. The annihilation probability is maximal for vanishing phase difference. This ``nonlocal Josephson effect'' is the superconducting analogue of the nonlocal Aharonov-Bohm effect \cite{Spl09} and the solid-state counterpart of the interferometry of superfluid Bose-Einstein condensates \cite{Iaz10}.

The spin-singlet pairing considered so far corresponds to a spin-$1/2$ Bogoliubov quasiparticle (electron and hole from opposite spin bands). The spin-up quasiparticle then annihilates with its spin-down counterpart. This is closest in analogy to the spin-$1/2$ Majorana fermion from particle physics (where neutrinos of opposite helicities would annihilate). In superconductors with strong spin-orbit coupling one can also consider a spinless Majorana fermion, with electron and hole from the same spin band (spin-triplet pairing). It is instructive to contrast the two cases. 

The transfer matrix for a spin-triplet Andreev edge channel has the form \eqref{Mpsinglet2} with a different electron-hole rotation matrix \cite{Ost11},
\begin{equation}
U(\alpha,\phi,\beta)=\exp\bigl[i\alpha(\tau_{x}\cos\phi+\tau_{y}\sin\phi)+i\beta\tau_{z}\bigr].\label{Utripletdef}
\end{equation}
The Pauli matrix $\sigma_{y}$ is no longer present, because electron and hole are from the same spin band. To preserve the particle-hole symmetry \eqref{Mehsym} the mixing strength $\alpha$ should be an odd function of energy: $\alpha(E)=-\alpha(-E)$. In particular, electron and hole are uncoupled at the Fermi energy ($E=0$). If we consider frequencies $\omega=2eV-\delta\omega$ near the upper cutoff, this energy dependence does not play a role (since the annihilating Bogoliubov quasiparticles then have the same energy $eV$). If we again take two identical superconductors we arrive at
\begin{equation}
P_{12}=-\tfrac{1}{4}G_{0} R(1-R)\delta\omega(1-\cos\phi_{12})\sin^{2}2\bar{\alpha}.\label{P12triplet}
\end{equation}
The factor of two difference with Eq.\ \eqref{P12simple} is due to the absence of spin degeneracy. The annihilation probability now vanishes for $\phi_{12}=0$. We interpret this in terms of Pauli blocking, operative because two Bogoliubov quasiparticles from the same spin band are indistinguishable for $\phi_{12}=0$. In the spin-singlet case, in contrast, they remain distinguished by their opposite spin.

To detect the nonlocal Josephson effect in an experiment, one would like to vary the superconducting phase difference $\phi_{12}$ without affecting the edge channels. This could be achieved by joining the two superconductors via a ring in the plane perpendicular to the 2D electron gas and then varying the flux through this ring. The resulting $h/2e$ oscillations in the noise correlator would have the largest amplitude for $\bar{\alpha}_1=\bar{\alpha}_2=\pi/4$, but there is no need for fine tuning of these parameters. For example, if only $\bar{\alpha}_1=\pi/4$, the amplitude of the oscillations varies as $\sin^2 2\alpha_2$, so it remains substantial for a broad interval of $\bar{\alpha}_2$ around $\pi/4$. 

The main experimental bottleneck is the coupling strength of the edge channel to the superconductor, which is of order $l_{\rm m}/l_{\rm S}$ (magnetic length over proximity-induced superconducting coherence length). The amplitude of the nonlocal Josephson oscillations depends quadratically on this ratio, so for $l_{\rm m}\simeq 10\,{\rm nm}$ (in a $4\,{\rm T}$ magnetic field) one would hope for a $l_{\rm S}$ below $100\,{\rm nm}$.  

In summary, we have proposed an experiment for Bogoliubov quasiparticles that is the condensed matter analogue of the way in which Majorana fermions are searched for in particle physics \cite{Bil12}: By detecting their pairwise annihilation upon collision. The Majorana fermions in a topologically trivial superconductor lack the non-Abelian statistics and the associated nonlocality of Majorana zero-modes in a topological superconductor \cite{Rea00}, but a different kind of nonlocality remains: We have found that the annihilation probability of quasiparticles originating from two identical superconductors depends on their phase difference --- even in the absence of any supercurrent coupling. Observation of the $h/2e$ oscillations of the annihilation probability would provide a striking demonstration of the Majorana nature of Bogoliubov quasiparticles.

I dedicate this paper to the memory of Markus B\"{u}ttiker. I have benefited from discussions with A. R. Akhmerov and from the support by the Foundation for Fundamental Research on Matter (FOM), the Netherlands Organization for Scientific Research (NWO/OCW), and an ERC Synergy Grant.

\appendix

\section{Response to feedback}

The following two appendices are in response to feedback on the manuscript that I received from Claudio Chamon and Yuli Nazarov.

\subsection{Symmetry of the current correlator}
\label{symmetry}

In some configurations the current correlator defined as in Eq.\ \eqref{Pomegadef},
\begin{equation}
P(\omega)=\int_{-\infty}^{\infty}dt\,e^{i\omega t}\langle I_{1}(0)I_{2}(t)\rangle,\label{Pomega1def}
\end{equation}
differs from the symmetrized version 
\begin{equation}
P_{\rm sym}(\omega)=\frac{1}{2}\int_{-\infty}^{\infty}dt\,e^{i\omega t}\langle  I_{1}(0)I_{2}(t)+I_{2}(t)I_{1}(0)\rangle.\label{Pomega2def}
\end{equation}
In our beam splitter configuration there is no difference: The two correlators are identical, because the current operators $I_{1}(t)$ and $I_{2}(t')$ commute. 

To see this, we start from the definition
\begin{align}
&I_{p}(t)=\frac{e}{4\pi}\int_{-\infty}^{\infty}dE\int_{-\infty}^{\infty}dE' \,e^{it(E-E')}\nonumber\\
&\qquad\quad\times c^{\dagger}(E){\cal M}_{p,z}(E,E')c(E'),\label{Ipdef}\\
&{\cal M}_{p,\alpha}(E,E')=S^{\dagger}(E){\cal P}_{p}\tau_{\alpha}S(E')={\cal M}^{\dagger}_{p,\alpha}(E',E),\label{calMdef}
\end{align}
where ${\cal P}_{p}$ projects onto contact $p$. The scattering matrix $S(E)$ relates quasiparticle operators before and after the beam splitter. It is a unitary matrix, constrained by particle-hole symmetry,
\begin{equation}
S(E)=\tau_{x}S^{\ast}(-E)\tau_{x}.\label{Ssymmetry}
\end{equation}
The fact that ${\cal P}_{p}{\cal P}_{q}=0$ if $p\neq q$ implies that
\begin{align}
&{\cal M}_{p,\alpha}(E,E'){\cal M}_{q,\beta}(E',E'')=0\;\;{\rm if}\;\;p\neq q.\label{MMrelation}\\
&{\cal M}_{p,\alpha}(E,E')\tau_{x}{\cal M}^{\rm T}_{q,\beta}(E'',-E')=0\;\;{\rm if}\;\;p\neq q,\label{MMrelation2}\\
&{\cal M}^{\rm T}_{p,\alpha}(-E,E')\tau_{x}{\cal M}_{q,\beta}(E,E'')=0\;\;{\rm if}\;\;p\neq q.\label{MMrelation3}
\end{align}

The Bogoliubov quasiparticle operators have the Majorana anticommutation relation, cf.\ Eq.\ \eqref{anticommutation}:
\begin{equation}
\begin{split}
&\{c(E),c^{\dagger}(E')\}=\delta(E-E')\tau_{0},\\
&\{c(E),c(E')\}=\delta(E+E')\tau_{x}.
\end{split}
\label{cccommutator}
\end{equation}
Substitution into the commutator $I_1(t)I_2(t')-I_2(t')I_1(t)$ produces four terms, which all vanish in view of Eqs.\ \eqref{MMrelation}--\eqref{MMrelation3}.

\subsection{Absence of supercurrent}
\label{supercurrent}

The dependence of the current correlator on the superconducting phases $\phi_{1},\phi_{2}$ is remarkable in view of the absence of any supercurrent coupling. The absence of supercurrent can be understood from Fig.\ \ref{fig_beamsplitter} by noting that the chirality of the edge states prevents the transfer of a Cooper pair between the two superconductors. More formally, one can calculate the density of states and ascertain that it is phase independent.

We use the relation
\begin{equation}
\rho(E)=(2\pi i)^{-1}\frac{d}{dE}\ln{\rm Det}\,S(E)\label{rhologDetS}
\end{equation}
between the density of states and the scattering matrix, which we construct from the scattering matrix $S_{\rm beam}$ of the beam splitter and the two transfer matrices $M_{1}$, $M_{2}$ of the edge states along the superconductors:
\begin{equation}
S(E)=S_{\rm beam}(E)\begin{pmatrix}
M_1(E)&0\\
0&M_2(E)
\end{pmatrix}.\label{SMrelation}
\end{equation}
The determinant factors into the product
\begin{equation}
{\rm Det}\,S={\rm Det}\,S_{\rm beam}\,{\rm Det}\,M_1\,{\rm Det}\,M_2.\label{DetSproduct} 
\end{equation}
The scattering matrix of the beam splitter is independent of $\phi_{p}$, while $M_{p}$ depends on $\phi_{p}$ according to Eq.\ \eqref{Mpsinglet2}:
\begin{align}
M_{p}&\propto \exp\bigl[i\bar{\alpha}_{p}\sigma_{y}\otimes(\tau_{x}\cos\phi_p+\tau_{y}\sin\phi_p)\bigr]\nonumber\\
&=\begin{pmatrix}
\cos\bar{\alpha}_p&ie^{-i\phi_p}\sigma_{y}\sin\bar{\alpha}_p\\
ie^{i\phi_p}\sigma_{y}\sin\bar{\alpha}_p&\cos\bar{\alpha}_p
\end{pmatrix}.\label{Mppropto}
\end{align}
The $\phi_p$-dependence drops out of ${\rm Det}\,M_{p}$, so ${\rm Det}\,S(E)$ and hence $\rho(E)$ are $\phi_p$-independent.


\begin{thebibliography}{99}
\bibitem{note2} In much of the literature the current operator \eqref{Itdef} of Bogoliubov quasiparticles is defined from the very beginning as an integral over only positive energies. This makes no difference for the electrical noise at zero frequency, but at nonzero frequencies it would miss the annihilation term $\propto\theta(\omega-eV)$ in Eq.\ \eqref{noisepower2}.
\end{thebibliography}
\end{document}